\documentclass[prd,nofootinbib]{revtex4}
\usepackage{amsmath}
\usepackage{graphicx}
\usepackage{amssymb}

\begin{document}

\title{Spinning compact binary inspiral: \\
Independent variables and dynamically preserved spin configurations}
\author{L\'{a}szl\'{o} \'{A}rp\'{a}d Gergely$^{1,2\star }$}
\affiliation{$^{1}$Department of Theoretical Physics, University of Szeged, Tisza Lajos
krt 84-86, Szeged 6720, Hungary\\
$^{2}$Department of Experimental Physics, University of Szeged, D\'{o}m t%
\'{e}r 9, Szeged 6720, Hungary\\
{\small {$^\star$ E-mail: gergely@physx.u-szeged.hu\qquad } }}

\begin{abstract}
We establish the set of independent variables suitable to monitor the
complicated evolution of the spinning compact binary during the inspiral.
Our approach is valid up to the second post-Newtonian order, including
leading order spin-orbit, spin-spin and mass quadrupol-mass monopole
effects, for generic (noncircular, nonspherical) orbits. Then we analyze the
conservative spin dynamics in terms of these variables. We prove that the
only binary black hole configuration allowing for spin precessions with
equal angular velocities about a common intantaneous axis roughly aligned
to the normal of the osculating orbit, is the \textit{equal mass and
parallel (aligned or antialigned) spin} configuration. This analytic result
puts limitations on what particular configurations can be selected in
numerical\ investigations of compact binary evolutions, even in those
including only the last orbits of the inspiral.
\end{abstract}

\date{\today }
\maketitle

\section{Introduction}

Compact objects are characterized by their size and gravitational radius
being comparable. They appear either as the end state of the stellar
evolution as neutron stars or black holes with a few solar masses (M$_{\odot
}$) or emerge from cosmological evolution by continued accretion and a
sequence of mergers \cite{MergerTree}\ as supermassive black holes of $%
3\times 10^{6}\div 3\times 10^{9}$ M$_{\odot }$, residing in the centers of
galaxies. Not much evidence has been gathered for the existence of
intermediate mass black holes (IMBH), although a detection of a variable
X-ray source of over 500 M$_{\odot }$ in the galaxy ESO 243-49 has been
recently reported and interpreted as IMBH \cite{IMBH Nature}. It has been
proposed that IMBHs ought to be searched for in globular clusters that can
be fitted well by medium-concentration King models \cite{IMBH Where}.

Compact objects are expected to frequently coexist in binary systems, formed
either by evolution of a stellar binary, by capture events or accompanying
the process of galaxy mergers. According to general relativity, compact
binaries radiate away gravitational waves, a process leading eventually to
their merger. Stellar mass binaries are among the most prominent sources for
the Earth-based gravitational wave detectors LIGO and Virgo \cite{LIGO},
while the gravitational waves produced during the (low mass) galactic black
hole mergers will be sought for by the long-planned space mission LISA \cite%
{LISA}.

The merging process can be split into three distinct phases. By definition
the inspiral is the regime of orbital evolution, which can be described
accurately in terms of a post-Newtonian (PN) expansion. Provided the orbits
are not excessively eccentric, the same PN parameter characterizes both weak
gravity and non-relativistic motion:%
\begin{equation}
\varepsilon =\frac{Gm}{c^{2}r}\approx \left( \frac{v}{c}\right) ^{2}~.
\label{PNparam}
\end{equation}%
A \textit{manifestly convergent} and finite procedure for calculating
gravitational radiation to arbitrary orders in a PN expansion was proposed 
\cite{convergenceWW}, based on solving a flat-spacetime wave equation
(representing Einstein equations with a harmonic gauge condition) as a
retarded integral over the past null cone of the chosen field point. A study
of the \textit{Cauchy convergence} for PN templates shows an oscillatory
behavior: increasing the PN order will not necessarily result in a better
template \cite{CauchyPN} (2PN templates being closer to numerical results,
than their 2.5 counterparts). The predictions of various PN approximants
(adiabatic Taylor, Pad\'{e} models, non-adiabatic effective-one-body models)
show that their convergence to numerical results is comparable \cite%
{convergenceTPEOB}. It is also known, that alternative template families
based on the shifted Chebyshev polynomials could exhibit faster Cauchy
convergence, than PN templates \cite{CauchyChebyshev}. Comparisons with full
general relativistic numerical runs confirmed that a third PN order approach
can be considered accurate for all practical purposes. The inspiral is
followed by the plunge, where a full general relativistic treatment is
necessary, and can be handled only numerically; and the ringdown, a process
during which all physical characteristics of the newly formed compact object
are radiated away, except mass, spin and possibly electric charge.

In this paper we investigate the conservative dynamics during the inspiral
of a spinning compact binary system. We include spin-orbit (SO), spin-spin
(SS) and mass quadrupole - mass monopole (QM) couplings, each to leading
order. The precession due to these interactions was first discussed in \cite%
{BOC}-\cite{BOC2}. With the spins and mass quadrupole moments included, the
number of variables in the configuration space increases considerably,
therefore we propose to find \textit{a minimal and conveniently chosen set
of independent variables}.

We note discussions of various aspects of the dynamics and gravitational
radiation related to the SO coupling in \cite{KWW}-\cite{Kidder}, SS
coupling in \cite{Kidder}-\cite{spinspin2}, and QM coupling in \cite{Poisson}%
-\cite{QM}. PN corrections to the SO coupling were presented in \cite{PNSO}
and the Hamiltonian approach including spins has been also worked out \cite%
{Ham}. Most recently, the back-reaction on the dynamics due to asymmetric
gravitational wave emission in the spinning case, possibly leading to strong
recoil effects, has been widely investigated, both analytically \cite{Kidder}%
, \cite{recoilSpinningAnalytical} and numerically for particular spin
configurations \cite{recoilSpinningNumerical}. Empirical formulae giving the
"final spin" have been advanced in Refs. \cite{finalspin} and some of them
compared in \cite{Barausse}. Zoom-whirl orbits (generic for particles
orbiting Kerr black holes \cite{zwKerr}) were also found in the framework of
the PN formalism \cite{zwPN}. A larger spin increases the likeliness of
apparition of such orbits \cite{zwNumeric}. Gravitational wave emission is
hold responsible for the occurrence of the spin-flip phenomenon \cite%
{LeahyWilliams}-\cite{Xspinflip} in X-shaped radio galaxies \cite%
{LeahyWilliams}, \cite{Xshape}. Recently it has been shown, that for a
typical merger of mass ratio at about $0.1$ the combined effect of SO
precession and gravitational radiation will result in the spin-flip
occurring during the inspiral \cite{SpinFlip}.

In Sec. \ref{KinDyn} we introduce the set of dynamical and configurational
variables characterizing the compact spinning binary. Both the
configurational and a subset of the dynamical variables depend on the choice
of the reference system. We use a number of four such systems, to be defined
in subsection \ref{RefSyst}, only one of them inertial, the rest of three
being rather adapted to the binary configuration. In subsection \ref%
{EulerConstr} we derive two relations among the time derivatives of the
introduced angular variables. As a result the time evolution of the
configurational variables is determined by the evolution of one single
configurational angle $\alpha $ and the true anomaly $\chi _{p}$. At the end
of this section we express the position and velocity vectors in the chosen
reference systems. As a by-product we recover the true anomaly
parametrization of the radial evolution, valid for the chosen perturbed
Keplerian setup.

Sec. \ref{genericSpin} introduces the angles characterizing the angular
momenta (total and orbital angular momenta and spins). The number of
independent variables characterizing them is shown to be 6. We will chose
them either as 5 angles and a scale, or equivalently as 3 angles and 3
scales.

In Sec. \ref{SpinEvol} we analyze the conservative evolution of the spins,
which is purely precessional, with the inclusion of the leading order
spin-orbit, spin-spin and mass quadrupole - mass monopole couplings. We
clarify the order (both PN and in the mass ratio) at which the various
contributions occur. Then we investigate, whether there are spin
configurations conserved by precessions, and we derive a no-go result.

The gravitational constant $G$ and speed of light $c$ are kept in all
expressions. For any vector $\mathbf{V}$ we denote its Euclidean magnitude
by $V$ and its direction by $\mathbf{\hat{V}}$.

\section{Kinematical and dynamical variables\label{KinDyn}}

\subsection{Variables}

We consider three distinct set of variables.

(a) \textit{The physical parameters of the binary}: The two compact objects
are characterized by masses $m_{i}$, spins $\mathbf{S}_{\mathbf{i}}$ ($i=1,2$%
), and mass quadrupole moments.

Equivalently to $m_{i}$ we can use the total mass $m\equiv m_{1}+m_{2}$ and
the reduced mass $\mu \equiv m_{1}m_{2}/m$. We assume that $m_{1}\geq m_{2}$%
. We also introduce the mass ratio $\nu \equiv m_{2}/m_{1}\leq 1$ and the
symmetric mass ratio $\eta \equiv \mu /m\in \left[ 0,0.25\right] $. The two
mass ratios are related as%
\begin{equation}
\eta =\frac{\nu }{\left( 1+\nu \right) ^{2}}~,  \label{etanu}
\end{equation}%
and for small $\nu $ we have $\eta =\allowbreak \nu -2\nu ^{2}+O\left( \nu
^{3}\right) $. We also note the useful relations%
\begin{equation}
m_{i}^{2}=m^{2}\eta \nu ^{2i-3}~.  \label{mi}
\end{equation}

Equivalently to $\mathbf{S}_{\mathbf{i}}$ we can introduce their magnitude,
polar and azimuthal angles. It is convenient to define dimensionless spin
magnitudes $\chi _{i}\in \left[ 0,1\right] $ by%
\begin{equation}
S_{i}\equiv \frac{G}{c}m_{i}^{2}\chi _{i}=\frac{G}{c}m^{2}\eta \nu
^{2i-3}\chi _{i}~.  \label{Si}
\end{equation}%
As for the spin angles, they depend on the chosen reference system. We will
discuss various possibilities in detail in Section \ref{genericSpin}.

We consider axisymmetric compact objects. Therefore the mass quadrupole of
the $i^{th}$ axially symmetric binary component is characterized by a single
quantity $Q_{i}$, its quadrupole-moment scalar \cite{Poisson}. Provided the
quadrupole moment originates entirely in its rotation (what we shall
assume), then the symmetry axis is $\mathbf{\hat{S}}_{i}$ and 
\begin{equation}
Q_{i}=-\frac{G^{2}}{c^{4}}w\chi _{i}^{2}m_{i}^{3}~,  \label{Qi}
\end{equation}%
with the parameter $w\in \left( 4,~8\right) $ for neutron stars, depending
on their equation of state, stiffer equations of state giving larger values
of $w$ \cite{Larakkers Poisson 1997}, \cite{Poisson}. For rotating black
holes $w=1$ \cite{Thorne 1980}. The negative sign arises because the
rotating compact object is centrifugally flattened, becoming an oblate
spheroid.

(b) \textit{Dynamical variables}: Up to 2PN accuracy the energy $E$ and the
total angular momentum vector $\mathbf{J\equiv L}+\mathbf{S}_{\mathbf{1}}+%
\mathbf{S}_{\mathbf{2}}$ are conserved \cite{KWW}. The orbital angular
momentum $\mathbf{L}$ and the spins $\mathbf{S}_{\mathbf{i}}$ are not
conserved separately, as the spins undergo a precessional motion. This will
be discussed in detail in Section \ref{SpinEvol}.

(c) \textit{Angular variables characterizing the orbit}:\qquad \qquad \qquad
\qquad \qquad \qquad \qquad \qquad \qquad \qquad \qquad \qquad \qquad \qquad
\qquad \qquad \qquad \qquad \qquad \qquad \qquad \qquad \qquad \qquad \qquad
\qquad \qquad \qquad \qquad \qquad \qquad \qquad

\begin{figure}[th]
\begin{center}
\includegraphics[height=10cm]{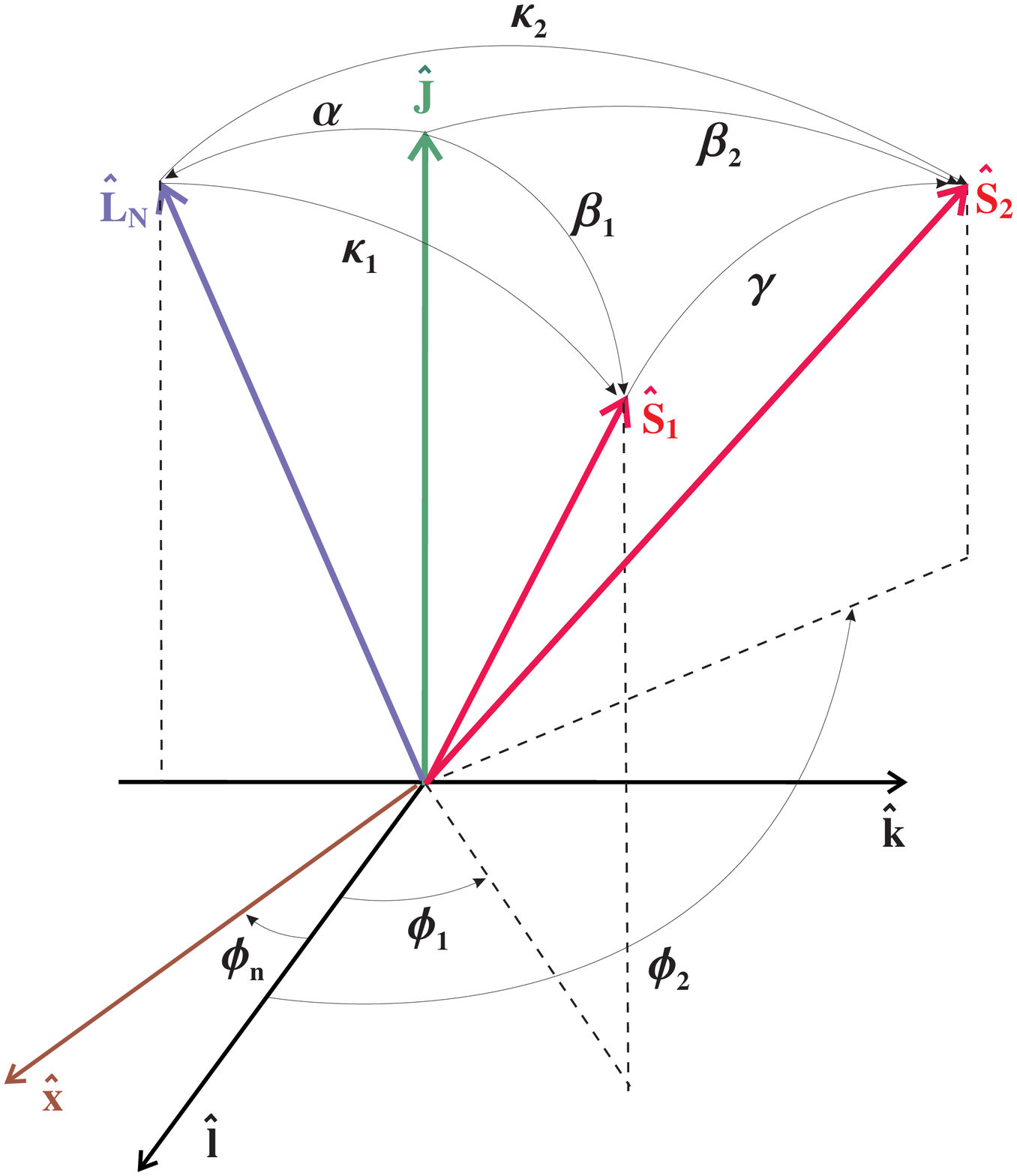}
\end{center}
\caption{The polar and azimuthal angles of the total angular momentum $J%
\mathbf{\hat{J}}$, Newtonian orbital angular momentum $L_{N}\mathbf{\hat{L}}%
_{\mathbf{N}}$ and spins $S_{1,2}\mathbf{\hat{S}}_{\mathbf{1,2}}$. Azimuthal
angles are shown in the non-inertial system $\mathcal{K}_{J}\equiv \left( 
\mathbf{\hat{l},\hat{k},\hat{J}}\right) $, polar angles both in $\mathcal{K}%
_{J}$ and respective to $\mathbf{\hat{L}}_{\mathbf{N}}$. The relative angle
of the spins is $\protect\gamma $. The non-inertial character of the system $%
\mathcal{K}_{J}$ is encoded in the evolution of the angle $\protect\phi _{n}$%
, measuring the angular separation of an inertial axis $\mathbf{\hat{x}}$
from the axis $\mathbf{\hat{l}}$.}
\label{fig1}
\end{figure}

\begin{figure}[th]
\begin{center}
\includegraphics[height=10cm]{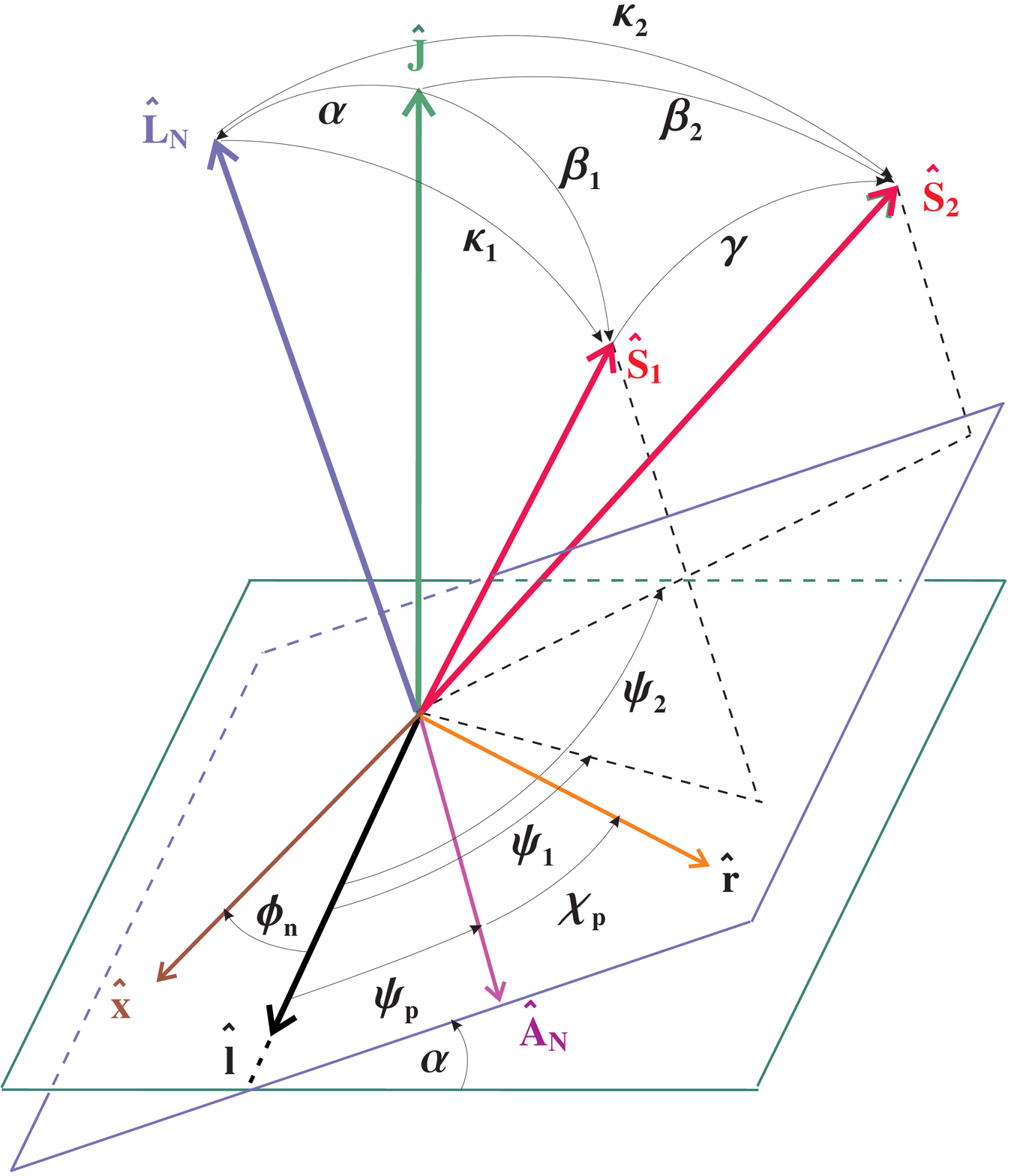}
\end{center}
\caption{The relative angles of the total angular momentum $J\mathbf{\hat{J}}
$, Newtonian orbital angular momentum $L_{N}\mathbf{\hat{L}}_{\mathbf{N}}$
and spins $S_{1,2}\mathbf{\hat{S}}_{\mathbf{1,2}}$ as in Fig \protect\ref%
{fig1}. The intersection of the planes perpendicular to $\mathbf{\hat{L}}_{%
\mathbf{N}}$ and $\mathbf{\hat{J}}$, respectively is the node line $\mathbf{%
\hat{l}}$. The inertial axis $\mathbf{\hat{x}}$ is at angle $\protect\phi %
_{n}$, measured from $\mathbf{\hat{l}}$ in the plane perpendicular to $%
\mathbf{\hat{J}}$. The azimuthal angles $\left( \protect\psi _{1},\protect%
\psi _{2},\protect\psi _{p}\right) $ of the spins and Newtonian
Laplace-Runge-Lenz vector $A_{N}\mathbf{\hat{A}}_{\mathbf{N}}$ (pointing
towards the periastron) are also measured from $\mathbf{\hat{l}}$, however
in the plane perpendicular to $\mathbf{\hat{L}}_{\mathbf{N}}$. The true
anomaly $\protect\chi _{p}$ is the angle between $\mathbf{\hat{A}}_{\mathbf{N%
}}$ and the position vector $r\mathbf{\hat{r}}$. Two of the basis vectors of
the inertial reference system $\mathcal{K}_{i}\equiv \left( \mathbf{\hat{x},%
\hat{y},\hat{J}}\right) $ and of each of the three non-inertial reference
systems $\mathcal{K}_{J}\equiv \left( \mathbf{\hat{l},\hat{k},\hat{J}}%
\right) $, $\mathcal{K}_{L}\equiv \left( \mathbf{\hat{l},\hat{m},\hat{L}}_{%
\mathbf{N}}\right) $, $\mathcal{K}_{A}\equiv \left( \mathbf{\hat{A}}_{%
\mathbf{N}}\mathbf{,\hat{Q}}_{\mathbf{N}}\mathbf{,\hat{L}}_{\mathbf{N}%
}\right) $ are shown. }
\label{fig2}
\end{figure}

The instantaneous orbital plane is perpendicular by definition to the
Newtonian orbital angular momentum $\mathbf{L_{N}}\equiv \mu \mathbf{r}%
\times \mathbf{v}$ and it evolves due to the spin precessions. We define (i)
the inclination $\alpha $ of the orbital plane with respect to the plane
perpendicular to $\mathbf{J}$ (thus $\alpha $~is the angle span by $\mathbf{%
\hat{L}}_{\mathbf{N}}$ and $\mathbf{\hat{J}}$); (ii) the angle $\phi _{n}$
between the intersection $\mathbf{\hat{l}}$ of these two planes and an
(arbitrary) inertial $x$-axis $\mathbf{\hat{x}}$ taken in the plane
perpendicular to $\mathbf{J}$, finally (iii) the angle $\psi _{p}$ measured
from $\mathbf{\hat{l}}$ to the periastron (see Figs \ref{fig1} and \ref{fig2}%
; the indices $p$ and $n$ stand for the periastron and node line,
respectively).

\subsection{Reference systems\label{RefSyst}}

For a better bookkeeping we introduce the inertial system $\mathcal{K}_{i}$
with $\mathbf{\hat{x}}$ and $\mathbf{\hat{J}}$ standing as the $x$- and $z$%
-axes and three non-inertial systems $\mathcal{K}_{J}$, $\mathcal{K}_{L}$
and $\mathcal{K}_{A}$.

In the system $\mathcal{K}_{J}$ the $z$-axis is fixed along $\mathbf{\hat{J}}
$, while in $\mathcal{K}_{L}$ along $\mathbf{\hat{L}}_{\mathbf{N}}$. We
choose $\mathbf{\hat{l}}=\mathbf{\hat{J}}\times \mathbf{\hat{L}}_{\mathbf{N}%
}/\sin \alpha $ as the $x$-axis of both systems. The system $\mathcal{K}_{J}$
is complete by $\mathbf{\hat{k}=\hat{J}}\times \mathbf{\hat{l}}$ and $%
\mathcal{K}_{L}$ by $\mathbf{\hat{m}=\hat{L}}_{\mathbf{N}}\times \mathbf{%
\hat{l}}$.

The system $\mathcal{K}_{A}$ also has $\mathbf{\hat{L}}_{\mathbf{N}}$ as the 
$z$-axis, however its $x$-axis is defined by the Laplace-Runge-Lenz vector 
\begin{equation}
\mathbf{A}_{\mathbf{N}}\equiv \mathbf{v}\times \mathbf{L}_{\mathbf{N}}-\frac{%
Gm\mu }{r}\mathbf{r}\ ,  \label{LRL}
\end{equation}%
which satisfies the constraints%
\begin{equation}
A_{N}^{2}=\frac{2E_{N}L_{N}^{2}}{\mu }+(Gm\mu )^{2}\ ,  \label{constr2}
\end{equation}%
and $\mathbf{L}_{\mathbf{N}}\mathbf{\cdot A}_{\mathbf{N}}=0$. Here $\mathbf{r%
}$ and $\mathbf{v}$ are the position vector and velocity of the reduced mass
particle orbiting $m$. The $y$-axis is defined by $\mathbf{Q}_{\mathbf{N}%
}\equiv \mathbf{L}_{\mathbf{N}}\times \mathbf{A}_{\mathbf{N}}$. The
orthonormal basis of $\mathcal{K}_{A}$ is therefore $(\mathbf{\hat{A}}_{%
\mathbf{N}},\ \mathbf{\hat{Q}}_{\mathbf{N}},~\mathbf{\hat{L}}_{\mathbf{N}})$.

The three angles $\left( \phi _{n},\alpha ,\psi _{p}\right) $ will be
referred to occasionally as Euler angles, as three consecutive rotations
with $-\phi _{n},~\alpha $ and $\psi _{p}$ about the axes $z,~x$ and again $%
z $ transform as $\mathcal{K}_{i}\rightarrow $ $\mathcal{K}_{J}\rightarrow $ 
$\mathcal{K}_{L}\rightarrow $ $\mathcal{K}_{A}$. The sequence of these
rotations is encompassed in the transformation matrix 
\begin{eqnarray}
\mathcal{R}\left( -\phi _{n},\alpha ,\psi _{p}\right) &=&\mathcal{R}%
_{z}\left( \psi _{p}\right) \mathcal{R}_{x}\left( \alpha \right) \mathcal{R}%
_{z}\left( -\phi _{n}\right)  \notag \\
&=&\left( 
\begin{array}{ccc}
\cos \psi _{p}\cos \phi _{n}+\sin \psi _{p}\cos \alpha \sin \phi _{n} & 
-\cos \psi _{p}\sin \phi _{n}+\sin \psi _{p}\cos \alpha \cos \phi _{n} & 
\sin \psi _{p}\sin \alpha \\ 
-\sin \psi _{p}\cos \phi _{n}+\cos \psi _{p}\cos \alpha \sin \phi _{n} & 
\sin \psi _{p}\sin \phi _{n}+\cos \psi _{p}\cos \alpha \cos \phi _{n} & \cos
\psi _{p}\sin \alpha \\ 
-\sin \alpha \sin \phi _{n} & -\sin \alpha \cos \phi _{n} & \cos \alpha%
\end{array}%
\right) ~,  \label{Euler}
\end{eqnarray}%
where $\mathcal{R}$ with one argument denotes the corresponding rotation
matrices acting on the coordinates.

\subsection{Constraints on the Euler angle evolutions\label{EulerConstr}}

The coordinates of the reduced mass particle in the inertial system $%
\mathcal{K}_{i}$ can be obtained by applying the transformation $\mathcal{R}%
\left( -\psi ,-\alpha ,\phi _{n}\right) $ to the coordinates of the vector $%
\mathbf{r}=r\left( 1,0,0\right) $. Here $\psi $ $=\psi _{p}+\chi _{p}$ is
the angle span by $\mathbf{\hat{l}}$ and $\mathbf{\hat{r}}$, with $\chi _{p}$
defined as the \textit{true anomaly}, the angle span by $\mathbf{\hat{A}}_{%
\mathbf{N}}$ and $\mathbf{\hat{r}}$. We obtain%
\begin{equation}
\left( 
\begin{array}{c}
x \\ 
y \\ 
z%
\end{array}%
\right) =r\left( 
\begin{array}{c}
\cos \phi _{n}\cos \psi +\sin \phi _{n}\cos \alpha \sin \psi \\ 
-\sin \phi _{n}\cos \psi +\cos \phi _{n}\cos \alpha \sin \psi \\ 
\sin \alpha \sin \psi%
\end{array}%
\right) ~.
\end{equation}%
A tedious, but straightforward computation carried on in the system $K_{i}$
gives%
\begin{eqnarray}
\frac{\mathbf{L_{N}}}{{\mu r}^{2}} &=&\dot{\phi}_{n}\left( 
\begin{array}{c}
\sin \alpha \sin \psi \left[ \cos \phi _{n}\cos \psi +\sin \phi _{n}\cos
\alpha \sin \psi \right] \\ 
\sin \alpha \sin \psi \left[ -\sin \phi _{n}\cos \psi +\cos \phi _{n}\cos
\alpha \sin \psi \right] \\ 
\sin ^{2}\alpha \sin ^{2}\psi -1%
\end{array}%
\right)  \notag \\
&&+\dot{\alpha}\sin \psi \left( 
\begin{array}{c}
-\sin \phi _{n}\cos \alpha \cos \psi +\cos \phi _{n}\sin \psi \\ 
-\cos \phi _{n}\cos \alpha \cos \psi -\sin \phi _{n}\sin \psi \\ 
-\sin \alpha \cos \psi%
\end{array}%
\right)  \notag \\
&&+\dot{\psi}\left( 
\begin{array}{c}
-\sin \phi _{n}\sin \alpha \\ 
-\cos \phi _{n}\sin \alpha \\ 
\cos \alpha%
\end{array}%
\right) ~.
\end{eqnarray}%
From here we readily obtain%
\begin{equation}
\frac{L_{N}^{2}}{{\mu }^{2}{r}^{4}}=\left( \dot{\psi}-\dot{\phi}_{n}\cos
\alpha \right) ^{2}+\left( \dot{\phi}_{n}\sin \alpha \cos \psi +\dot{\alpha}%
\sin \psi \right) ^{2}~.  \label{LN2dots}
\end{equation}%
Also, dividing the third component (which by definition is $\left( \mathbf{%
L_{N}}\right) _{z}=L_{N}\cos \alpha $) by $\cos \alpha $ we get 
\begin{equation}
\frac{L_{N}}{{\mu r}^{2}}=\dot{\psi}-\dot{\phi}_{n}\cos \alpha -\left( \dot{%
\phi}_{n}\sin \alpha \cos \psi +\dot{\alpha}\sin \psi \right) \tan \alpha
\cos \psi ~.  \label{LNzdots}
\end{equation}%
In the Newtonian approximation the Euler angles being constant, we recover $%
L_{N}=\mu r^{2}\dot{\chi}_{p}$ and $\left( \mathbf{L_{N}}\right) _{z}=\mu
r^{2}\dot{\chi}_{p}\cos \alpha $.

By squaring Eq. (\ref{LNzdots}) and subtracting from Eq. (\ref{LN2dots}) we
obtain the identity:%
\begin{eqnarray}
0 &=&\left[ \left( 1-\tan ^{2}\alpha \cos ^{2}\psi \right) \left( \dot{\phi}%
_{n}\sin \alpha \cos \psi +\dot{\alpha}\sin \psi \right) +2\tan \alpha \cos
\psi \left( \dot{\psi}-\dot{\phi}_{n}\cos \alpha \right) \right]  \notag \\
&&\times \left( \dot{\phi}_{n}\sin \alpha \cos \psi +\dot{\alpha}\sin \psi
\right) ~.
\end{eqnarray}%
The first factor cannot vanish, as to Newtonian order it gives $2\tan \alpha
\cos \psi L_{N}/\mu r^{2}\neq 0$, therefore the vanishing of the second
factor (by reintroducing $\psi =\psi _{p}+\chi _{p}$) gives:%
\begin{equation}
\dot{\phi}_{n}=-\dot{\alpha}\frac{\tan \left( \psi _{p}+\chi _{p}\right) }{%
\sin \alpha }~.  \label{firstEq}
\end{equation}%
Reinserting this in either of the Eqs. (\ref{LN2dots}) or (\ref{LNzdots})
gives 
\begin{equation}
\dot{\psi}_{p}+\dot{\chi}_{p}=\frac{L_{N}}{{\mu r}^{2}}+\dot{\phi}_{n}\cos
\alpha ~.  \label{secondEq}
\end{equation}%
We have just derived two relations among the time derivatives of the Euler
angles and of the true anomaly, which restrict the number of independent
angular variables introduced up to now to $\alpha $ and $\chi _{p}$.

\subsection{The position and velocity vectors in the bases $\mathcal{K}_{A}$
and $\mathcal{K}_{L}$}

Simple computation starting from the definitions of $\mathbf{A}_{\mathbf{N}}$
and $\mathbf{Q}_{\mathbf{N}}$ gives 
\begin{eqnarray}
\mathbf{A}_{\mathbf{N}} &=&\mu \left( \frac{2E_{N}}{\mu }+\frac{Gm}{r}%
\right) \mathbf{r-}\mu r\dot{r}\mathbf{v}\ ,  \notag \\
\mathbf{Q}_{\mathbf{N}} &=&Gm\mu ^{2}\dot{r}\mathbf{r+}\left(
L_{N}^{2}-Gm\mu ^{2}r\right) \mathbf{v}\ .  \label{basisnew}
\end{eqnarray}%
From here the expressions of the position and velocity vectors in $\mathcal{K%
}_{A}$ emerge as%
\begin{eqnarray}
\mathbf{r} &=&\frac{L_{N}^{2}-Gm\mu ^{2}r}{\mu A_{N}}\mathbf{\hat{A}}_{%
\mathbf{N}}+\frac{L_{N}}{A_{N}}r\dot{r}\mathbf{\hat{Q}}_{\mathbf{N}}\ ,
\label{rv} \\
\mathbf{v} &=&-\frac{Gm\mu }{A_{N}}\dot{r}\mathbf{\hat{A}}_{\mathbf{N}}+%
\frac{L_{N}}{A_{N}}\left( \frac{2E_{N}}{\mu }+\frac{Gm}{r}\right) \mathbf{%
\hat{Q}}_{\mathbf{N}}\ .  \label{vv}
\end{eqnarray}%
In terms of the true anomaly $\chi _{p}$ (the azimuthal angle of $\mathbf{r}$
in the system $\mathcal{K}_{A}$), the position vector is given by%
\begin{equation}
\mathbf{r}=r\left( \cos \chi _{p}\mathbf{\hat{A}}_{\mathbf{N}}+\sin \chi _{p}%
\mathbf{\hat{Q}}_{\mathbf{N}}\right) \mathbf{~,}  \label{rKA}
\end{equation}%
which compared with Eq. (\ref{rv}) gives \textit{the true anomaly
parametrization}:%
\begin{eqnarray}
r &=&\frac{L_{N}^{2}}{\mu \left( Gm\mu +A_{N}\cos \chi _{p}\right) }~,
\label{truer} \\
\dot{r} &=&\frac{A_{N}}{L_{N}}\sin \chi _{p}~.  \label{truerdot}
\end{eqnarray}%
In terms of the true anomaly, the velocity is expressed as%
\begin{equation}
\mathbf{v}=\frac{Gm\mu }{L_{N}}\left[ -\sin \chi _{p}\mathbf{\hat{A}}_{%
\mathbf{N}}+\left( \cos \chi _{p}+\frac{A_{N}}{Gm\mu }\right) \mathbf{\hat{Q}%
}_{\mathbf{N}}\right] \ .  \label{vKA}
\end{equation}%
Its square gives $v^{2}$ in terms of the true anomaly:%
\begin{equation}
v^{2}=\frac{\left( Gm\mu \right) ^{2}+A_{N}^{2}+2Gm\mu A_{N}\cos \chi _{p}}{%
L_{N}^{2}}~.  \label{truev2}
\end{equation}%
(The same emerges from the definition of the Newtonian energy $E_{N}\equiv
\mu v^{2}/2-Gm\mu /r$, by applying Eqs. (\ref{constr2}) and (\ref{truer}).)

As the basis vectors of $\mathcal{K}_{A}$ are related to the basis vectors
of $\mathcal{K}_{L}$ by a rotation with angle $\psi _{p}$: 
\begin{eqnarray}
\mathbf{\hat{A}}_{\mathbf{N}} &=&\cos \psi _{p}\mathbf{\hat{l}}+\sin \psi
_{p}\mathbf{\hat{m}~,}  \notag \\
\mathbf{\hat{Q}}_{\mathbf{N}} &=&-\sin \psi _{p}\mathbf{\hat{l}}+\cos \psi
_{p}\mathbf{\hat{m}~,}  \label{KLA}
\end{eqnarray}%
it is straightforward to rewrite $\mathbf{r}$ and $\mathbf{v}$\ in the basis 
$\mathcal{K}_{L}$ as 
\begin{eqnarray}
\mathbf{r} &=&r\left( \cos \psi \mathbf{\hat{l}}+\sin \psi \mathbf{\hat{m}}%
\right) \ ,  \label{rKL} \\
\mathbf{v} &=&\frac{1}{L_{N}}\left[ -\left( Gm\mu \sin \psi +A_{N}\sin \psi
_{p}\right) \mathbf{\hat{l}}+\left( Gm\mu \cos \psi +A_{N}\cos \psi
_{p}\right) \mathbf{\hat{m}}\right] \ .  \label{vKL}
\end{eqnarray}

\section{Constraints on angular momentum variables\label{genericSpin}}

\subsection{The 5 angular degrees of freedom}

The polar and azimuthal angles of $\mathbf{\hat{L}}_{\mathbf{N}}$ and $%
\mathbf{\hat{S}_{i}}$ in $\mathcal{K}_{J}$ are ($\alpha ,-\pi /2$) and ($%
\beta _{i},\phi _{i}$), respectively, such that%
\begin{eqnarray}
\mathbf{\hat{L}}_{\mathbf{N}} &=&-\sin \alpha \mathbf{\hat{k}}+\cos \alpha 
\mathbf{\hat{J}}~,  \label{LdirKJ} \\
\mathbf{\hat{S}_{i}} &=&\sin \beta _{i}\cos \phi _{i}\mathbf{\hat{l}}+\sin
\beta _{i}\sin \phi _{i}\mathbf{\hat{k}}+\cos \beta _{i}\mathbf{\hat{J}}~.
\label{SidirKJ}
\end{eqnarray}%
Similarly, the polar and azimuthal angles of $\mathbf{\hat{J}}$ and $\mathbf{%
\hat{S}_{i}}$ in $\mathcal{K}_{L}$ are ($\alpha ,\pi /2$) and ($\kappa
_{i},\psi _{i}$), respectively, thus%
\begin{eqnarray}
\mathbf{\hat{J}} &=&\sin \alpha \mathbf{\hat{m}}+\cos \alpha \mathbf{\hat{L}}%
_{\mathbf{N}}~,  \label{JdirKL} \\
\mathbf{\hat{S}_{i}} &=&\sin \kappa _{i}\cos \psi _{i}\mathbf{\hat{l}}+\sin
\kappa _{i}\sin \psi _{i}\mathbf{\hat{m}+}\cos \kappa _{i}\mathbf{\hat{L}}_{%
\mathbf{N}}~.  \label{SidirKL}
\end{eqnarray}%
By comparing the two forms of the $\mathbf{\hat{l}}$ component of the
vectors $\mathbf{\hat{S}_{i}}$ we get%
\begin{equation}
\sin \kappa _{i}\cos \psi _{i}=\sin \beta _{i}\cos \phi _{i}~.
\label{kapsibephi}
\end{equation}%
By computing $\mathbf{\hat{S}_{i}\cdot \hat{L}}_{\mathbf{N}}$ in both
systems we find 
\begin{equation}
\cos \kappa _{i}\mathbf{=}\cos \alpha \cos \beta _{i}-\sin \alpha \sin \beta
_{i}\sin \phi _{i}~.  \label{sphe1}
\end{equation}%
As $\sin \phi _{i}=-\cos \left( \pi /2+\phi _{i}\right) $ and $\pi /2+\phi
_{i}$ is the relative azimuthal angle of $\mathbf{\hat{L}}_{\mathbf{N}}$ and 
$\mathbf{\hat{S}_{i}}$, Eq. (\ref{sphe1}) is but the spherical cosine
identity in the triangle defined by these three vectors on the unit sphere.

Similarly, from the two expressions $\mathbf{\hat{S}_{1}\cdot \hat{S}_{2}}%
\equiv \cos \gamma $ written in both reference systems we find the spherical
cosine identities:%
\begin{eqnarray}
\cos \gamma &=&\cos \kappa _{1}\cos \kappa _{2}+\sin \kappa _{1}\sin \kappa
_{2}\cos \Delta \psi ~,  \label{gombi} \\
\cos \gamma &=&\cos \beta _{1}\cos \beta _{2}+\sin \beta _{1}\sin \beta
_{2}\cos \Delta \phi ~.  \label{gombi1}
\end{eqnarray}%
where $\Delta \psi =\psi _{2}-\psi _{1}$ and $\Delta \phi =\phi _{2}-\phi
_{1}$ are the differences in the azimuthal angles of the two spins in the
two systems $\mathcal{K}_{L}$ and $\mathcal{K}_{J}$, respectively.

Other spherical triangle identities arise by computing $\mathbf{\hat{S}%
_{i}\cdot \hat{J}}$ in both systems: 
\begin{equation}
\cos \beta _{i}=\cos \alpha \cos \kappa _{i}+\sin \alpha \sin \kappa
_{i}\sin \psi _{i}~.  \label{sphe}
\end{equation}%
Then Eqs. (\ref{kapsibephi}) and (\ref{sphe}) give $\beta _{i},\phi _{i}$ as
function of $\kappa _{i},\psi _{i}$ and $\alpha $. Inserting these in Eqs. (%
\ref{gombi}) and (\ref{gombi1}) and eliminating $\gamma $ could in principle
give $\alpha $ as function of $\kappa _{i},\psi _{i}$ alone. We get:%
\begin{equation}
\sin \beta _{1}\sin \phi _{1}\sin \beta _{2}\sin \phi _{2}=\left( \sin
\alpha \cos \kappa _{1}-\cos \alpha \sin \kappa _{1}\sin \psi _{1}\right)
\left( \sin \alpha \cos \kappa _{2}-\cos \alpha \sin \kappa _{2}\sin \psi
_{2}\right) ~.
\end{equation}%
As the orientation of the spins are independent, we obtain%
\begin{equation}
\sin \beta _{i}\sin \phi _{i}=\sin \alpha \cos \kappa _{i}-\cos \alpha \sin
\kappa _{i}\sin \psi _{i}~,
\end{equation}%
however the direct computation of the left hand side by employing Eqs. (\ref%
{kapsibephi}) and (\ref{sphe}) results in the right hand side, leading to an
identity rather than an expression of $\alpha $ as function of $\kappa
_{i},\psi _{i}$. Therefore Eq. (\ref{gombi1}) can be considered as a
consequence of the other equations. Similarly one can show that Eqs. (\ref%
{sphe1}) are consequences of the other equations.

We conclude that there are 5 independent constraint equations for the 10
angles ($\alpha ,\beta _{i},\phi _{i},\kappa _{i},\psi _{i},\gamma $),
namely Eqs. (\ref{kapsibephi}), (\ref{sphe}) and (\ref{gombi}), and we can
take ($\alpha ,\kappa _{i},\psi _{i}$) as the independent angles. The
network of all angles in the systems $K_{J}$ and $K_{L}$ is represented on
Figs \ref{fig1} and \ref{fig2}, respectively.

\subsection{Orbital angular momentum}

The total orbital angular momentum $\mathbf{L}$ contains pure general
relativistic (PN, 2PN) and spin-orbit (SO) contributions \cite{Kidder}:%
\footnote{%
The equations of motion leading to this expression were derived in harmonic
coordinates, imposing the covariant spin supplementary condition.}%
\begin{equation}
\mathbf{L}=\mathbf{L_{N}}+\mathbf{L_{PN}}+\mathbf{L_{SO}+L_{2PN}}\ .
\label{Lvect}
\end{equation}%
There are no spin-spin or quadrupole-monopole contributions to the orbital
angular momentum \cite{quadrup}. Here the $\mathbf{L_{PN}}$ and $\mathbf{%
L_{2PN}}$ contributions are aligned to $\mathbf{L_{N}}$ (cf. Eq. (2.9) of
Ref. \cite{Kidder}):%
\begin{eqnarray}
\mathbf{L}_{\mathbf{PN}} &=&\epsilon _{PN}\mathbf{L}_{\mathbf{N}}\ ,  \notag
\\
\epsilon _{PN} &=&\frac{1-3\eta }{2}\left( \frac{v}{c}\right) ^{2}+\left(
3+\eta \right) \frac{Gm}{c^{2}r}~,  \label{epPN}
\end{eqnarray}%
and 
\begin{eqnarray}
\mathbf{L}_{2\mathbf{PN}} &=&\epsilon _{2PN}\mathbf{L}_{\mathbf{N}}\ , 
\notag \\
\epsilon _{2PN} &=&\frac{3}{8}\left( 1-7\eta +13\eta ^{2}\right) \left( 
\frac{v}{c}\right) ^{4}-\frac{1}{2}\eta \left( 2+5\eta \right) \frac{Gm}{%
c^{2}r}\left( \frac{\dot{r}}{c}\right) ^{2}  \notag \\
&&+\frac{1}{2}\left( 7-10\eta -9\eta ^{2}\right) \frac{Gm}{c^{2}r}\left( 
\frac{v}{c}\right) ^{2}+\frac{1}{4}\left( 14-41\eta +4\eta ^{2}\right)
\left( \frac{Gm}{c^{2}r}\right) ^{2}~.  \label{ep2PN}
\end{eqnarray}%
The SO contribution (Eq. (2.9.c) of Ref. \cite{Kidder}) can be rewritten as 
\begin{eqnarray}
\mathbf{L_{SO}} &=&\sum_{i=1}^{2}S_{i}\left[ \epsilon _{\mathbf{i}}^{\mathbf{%
r}}\left( \mathbf{\hat{r}}\cdot \mathbf{\hat{S}_{i}}\right) \mathbf{\hat{r}}%
+\epsilon _{\mathbf{i}}^{\mathbf{v}}\left( \mathbf{\hat{v}}\cdot \mathbf{%
\hat{S}_{i}}\right) \mathbf{\hat{v}-}\left( \epsilon _{\mathbf{i}}^{\mathbf{r%
}}+\epsilon _{\mathbf{i}}^{\mathbf{v}}\right) \mathbf{\hat{S}_{i}}\right] ~,
\notag \\
\epsilon _{\mathbf{i}}^{\mathbf{r}} &=&{\frac{Gm}{c^{2}r}\eta }\left( 2+\nu
^{3-2i}\right)   \notag \\
\epsilon _{\mathbf{i}}^{\mathbf{v}} &=&-{\frac{v^{2}}{2c^{2}}\eta }\nu
^{3-2i}~.  \label{epSO}
\end{eqnarray}%
Note that 
\begin{eqnarray}
\epsilon _{PN} &=&\mathcal{O}\left( \varepsilon \right) \mathcal{O}\left(
1,\eta \right) ~,  \notag \\
\epsilon _{2PN} &=&\mathcal{O}\left( \varepsilon ^{2}\right) \mathcal{O}%
\left( 1,\eta ,\eta ^{2}\right) ~,  \label{epordo1}
\end{eqnarray}%
and 
\begin{eqnarray}
\epsilon _{\mathbf{i}}^{\mathbf{r}} &=&\mathcal{O}\left( \varepsilon \right) 
\mathcal{O}\left( \eta \right) \mathcal{O}\left( 1,\nu ^{3-2i}\right) ~, 
\notag \\
\epsilon _{\mathbf{i}}^{\mathbf{v}} &=&\mathcal{O}\left( \varepsilon \right) 
\mathcal{O}\left( \eta \right) \mathcal{O}\left( \nu ^{3-2i}\right) ~.
\label{epordo2}
\end{eqnarray}%
In order to evaluate the PN\ order of the $\mathbf{L_{SO}}$ contribution in $%
\mathbf{J}$, we evaluate on circular orbits%
\begin{eqnarray}
\frac{S_{i}}{L_{N}} &=&\frac{\left( G/c\right) m^{2}\eta \nu ^{2i-3}\chi _{i}%
}{\mu rv}=\left( \frac{Gm}{c^{2}r}\right) \frac{c}{v}\nu ^{2i-3}\chi _{i} 
\notag \\
&=&\mathcal{O}\left( \varepsilon ^{1/2}\right) \mathcal{O}\left( \nu
^{2i-3}\right) \chi _{i}~,  \label{SiLNordo2}
\end{eqnarray}%
which continue to approximately hold for eccentric orbits. This reasoning
shows that the SO contribution is of 1.5 PN order and also indicates how to
pick up the dominant terms when the mass ratio is small or when one would
like to employ a less accurate, but simpler description, dropping higher
order terms.

The total angular momentum is then%
\begin{equation}
J\mathbf{\hat{J}}=\left( 1+\epsilon _{PN}+\epsilon _{2PN}\right) L_{N}%
\mathbf{\hat{L}}_{\mathbf{N}}+\sum_{i=1}^{2}S_{i}\left[ \epsilon _{\mathbf{i}%
}^{\mathbf{r}}\left( \mathbf{\hat{r}}\cdot \mathbf{\hat{S}_{i}}\right) 
\mathbf{\hat{r}}+\epsilon _{\mathbf{i}}^{\mathbf{v}}\left( \mathbf{\hat{v}}%
\cdot \mathbf{\hat{S}_{i}}\right) \mathbf{\hat{v}+}\left( 1\mathbf{-}%
\epsilon _{\mathbf{i}}^{\mathbf{r}}-\epsilon _{\mathbf{i}}^{\mathbf{v}%
}\right) \mathbf{\hat{S}_{i}}\right] \ .  \label{Jdirrv}
\end{equation}

\subsection{One scaling degree of freedom}

In this subsection we will employ the projections of the Eq. \textbf{(}\ref%
{Jdirrv}\textbf{) }in order to derive relations between the angles and
magnitudes of the angular momenta involved. In the $\mathcal{K}_{L}$ system
the projections along the axes $\mathbf{\hat{l}}$, $\mathbf{\hat{m}}$ and $%
\mathbf{\hat{L}}_{\mathbf{N}}$ give, respectively:%
\begin{eqnarray}
0 &=&\sum_{i=1}^{2}S_{i}\sin \kappa _{i}\left\{ \left( 1\mathbf{-}\epsilon _{%
\mathbf{i}}^{\mathbf{r}}-\epsilon _{\mathbf{i}}^{\mathbf{v}}\right) \cos
\psi _{i}+\epsilon _{\mathbf{i}}^{\mathbf{r}}\cos \left( \psi -\psi
_{i}\right) \cos \psi +\epsilon _{\mathbf{i}}^{\mathbf{v}}\mathcal{S}\left[
Gm\mu \sin \left( \psi -\psi _{i}\right) +A_{N}\sin \left( \psi _{p}-\psi
_{i}\right) \right] \right\} \ ,  \label{proj1} \\
J\sin \alpha &=&\sum_{i=1}^{2}S_{i}\sin \kappa _{i}\left\{ \left( 1\mathbf{-}%
\epsilon _{\mathbf{i}}^{\mathbf{r}}-\epsilon _{\mathbf{i}}^{\mathbf{v}%
}\right) \sin \psi _{i}+\epsilon _{\mathbf{i}}^{\mathbf{r}}\cos \left( \psi
-\psi _{i}\right) \sin \psi -\epsilon _{\mathbf{i}}^{\mathbf{v}}\mathcal{C}%
\left[ Gm\mu \sin \left( \psi -\psi _{i}\right) +A_{N}\sin \left( \psi
_{p}-\psi _{i}\right) \right] \right\} \ ,  \label{proj2} \\
J\cos \alpha &=&L_{N}\left( 1+\epsilon _{PN}+\epsilon _{2PN}\right)
+\sum_{i=1}^{2}S_{i}\left( 1\mathbf{-}\epsilon _{\mathbf{i}}^{\mathbf{r}%
}-\epsilon _{\mathbf{i}}^{\mathbf{v}}\right) \cos \kappa _{i}\ ,
\label{proj3}
\end{eqnarray}%
where%
\begin{eqnarray}
\mathcal{S}\left( \chi _{p},\psi _{p}\right) &=&\frac{Gm\mu \sin \left( \psi
_{p}+\chi _{p}\right) +A_{N}\sin \psi _{p}}{\left( Gm\mu \right)
^{2}+A_{N}^{2}+2Gm\mu A_{N}\cos \chi _{p}}~,  \notag \\
\mathcal{C}\left( \chi _{p},\psi _{p}\right) &=&\frac{Gm\mu \cos \left( \psi
_{p}+\chi _{p}\right) +A_{N}\cos \psi _{p}}{\left( Gm\mu \right)
^{2}+A_{N}^{2}+2Gm\mu A_{N}\cos \chi _{p}}~.
\end{eqnarray}%
In the derivation of Eqs. (\ref{proj1})-(\ref{proj3}) we have employed Eqs. (%
\ref{rKL}), (\ref{vKL}), (\ref{JdirKL}), (\ref{SidirKL}) from where we also
obtained 
\begin{eqnarray}
\mathbf{\hat{r}}\cdot \mathbf{\hat{S}_{i}} &=&\sin \kappa _{i}\cos \left(
\psi -\psi _{i}\right) ~,  \label{rS} \\
\mathbf{\hat{v}}\cdot \mathbf{\hat{S}_{i}} &\mathbf{=}&\frac{\sin \kappa _{i}%
}{L_{N}v}\left[ -Gm\mu \sin \left( \psi -\psi _{i}\right) +A_{N}\sin \left(
\psi _{i}-\psi _{p}\right) \right] ~,  \label{vS}
\end{eqnarray}%
with $v$ given by Eq. (\ref{truev2}).

We thus have introduced the 14 quantities ($J,L,\chi _{i},\alpha ,\beta
_{i},\phi _{i},\kappa _{i},\psi _{i},\gamma $) describing the angular
momenta,\ which are constrained by 8 independent \ relations. This leaves us
with 6 independent variables. 5 of these can be thought as the angles
defining the directions of the spins and orbital angular momentum in the $%
K_{L}$ system ($\alpha ,\kappa _{i},\psi _{i}$), a sixth one being a linear
scale, most conveniently chosen as $J$.

Note that in Eqs. (\ref{proj1})-(\ref{proj3}) the coefficients $\epsilon
_{PN}$, $\epsilon _{2PN}$, $\epsilon _{\mathbf{i}}^{\mathbf{r}}$, $\epsilon
_{\mathbf{i}}^{\mathbf{v}}$ depend only on the masses and $\chi _{p}$.
Therefore all dependences on $\psi _{i}$ are explicit. In principle Eqs. (%
\ref{proj1})-(\ref{proj2}) can be used to express $\psi _{i}$ as function of 
$\kappa _{i}$, $\alpha $, $\psi _{p}$, the masses and the spins $\chi _{i}$.
In practice however this may be cumbersome. The easiest way to do it is to
rewrite both the $\sin \psi _{i}$ and $\cos \psi _{i}$ in terms of the
variables $x_{i}=\tan \psi _{i}/2$. Then Eqs. (\ref{proj1})-(\ref{proj2})
become second rank coupled polynomial equations, possibly leading to two
distinct values of $\psi _{i}$ for each $\chi _{i}$.

Finally, Eq. (\ref{proj3}) can be employed to eliminate $L_{N}$ in the
detriment of the angular variables, spins and $J$, by a series expansion in $%
\varepsilon $ to 2PN order accuracy as%
\begin{eqnarray}
L_{N} &=&J\left( 1-\epsilon _{PN}+\epsilon _{PN}^{2}-\epsilon _{2PN}\right)
\cos \alpha -\sum_{i=1}^{2}S_{i}\left( 1-\epsilon _{PN}-\epsilon _{\mathbf{i}%
}^{\mathbf{r}}-\epsilon _{\mathbf{i}}^{\mathbf{v}}+\epsilon _{PN}\epsilon _{%
\mathbf{i}}^{\mathbf{r}}+\epsilon _{PN}\epsilon _{\mathbf{i}}^{\mathbf{v}%
}+\epsilon _{PN}^{2}-\epsilon _{2PN}\right) \cos \kappa _{i}  \notag \\
&=&\left( 1-\epsilon _{PN}+\epsilon _{PN}^{2}-\epsilon _{2PN}\right)
L_{N,0}+\sum_{i=1}^{2}S_{i}\left( 1-\epsilon _{PN}\right) \left( \epsilon _{%
\mathbf{i}}^{\mathbf{r}}+\epsilon _{\mathbf{i}}^{\mathbf{v}}\right) \cos
\kappa _{i}\ ,  \label{LNtaylor}
\end{eqnarray}%
where 
\begin{equation}
L_{N,0}=J\cos \alpha -\sum_{i=1}^{2}S_{i}\cos \kappa _{i}  \label{LN0}
\end{equation}%
is the leading order contribution to the orbital angular momentum, arising
when we approximate $\mathbf{J}$ as the sum of the Newtonian orbital angular
momentum and the spins. For convenience we also give%
\begin{equation}
\frac{1}{L_{N}}=\frac{1+\epsilon _{PN}+\epsilon _{2PN}}{L_{N,0}}-\frac{%
\sum_{i=1}^{2}S_{i}\left( 1+\epsilon _{PN}\right) \left( \epsilon _{\mathbf{i%
}}^{\mathbf{r}}+\epsilon _{\mathbf{i}}^{\mathbf{v}}\right) \cos \kappa _{i}}{%
L_{N,0}^{2}}~.
\end{equation}

\subsection{Summary: the independent variables}

The considerations in this section leave us with the following alternative
sets of independent variables, all characterizing the angular momenta: $%
\left( \alpha ,\kappa _{i},\psi _{i},J\right) $ or $\left( \alpha ,\kappa
_{i},\chi _{i},J\right) $. The second set represents the most advantageous
way of choosing the variables. Most notably, while $\psi _{i}$ are constant
over the orbital scale, they vary with the precessions. By contrast $\chi
_{i}$ are constant over the precession time-scale either, moreover they are
unaffected by gravitational radiation reaction, to quite high PN orders.
Also, $J$ stays constant up to 2PN accuracy (thus over precession
time-scale) as opposed to either of $L$, $L_{N}$, $L_{N,0}$. It changes only
over the radiation time-scale.

Once the evolution of $\chi _{p}$ is known, the other two Euler angles $%
\left( \phi _{n},\psi _{p}\right) $ become determined by the rest of
variables through Eqs. (\ref{firstEq}) and (\ref{secondEq}).

\section{Spin evolution\label{SpinEvol}}

The spins obey a precessional motion, as was derived for bodies with
arbitrary, but constant mass, spin and quadrupole moments (Eqs. (39) and
(43) of Ref. \cite{BOC}, see also Ref. \cite{BOC2}):%
\begin{equation}
\mathbf{\dot{S}_{i}}=\mathbf{\Omega }_{\mathbf{i}}\times \mathbf{S_{i}}\ ,
\label{Sprec}
\end{equation}%
with the angular velocities consisting of SO, SS and QM contributions. The
latter come from regarding each of the binary components as a mass monopole
moving in the quadrupolar field of the other component.

The precessional angular velocity is decomposed as 
\begin{eqnarray}
\mathbf{\Omega }_{\mathbf{i}} &=&\mathbf{\Omega }_{\mathbf{i}}^{SO}+\mathbf{%
\Omega }_{\mathbf{i}}^{SS}+\mathbf{\Omega }_{\mathbf{i}}^{QM}~,  \notag \\
\mathbf{\Omega }_{\mathbf{i}}^{SO} &=&{\frac{G\left( 4+3\nu ^{3-2i}\right) }{%
2c^{2}r^{3}}L}_{N}\mathbf{\hat{L}_{N}}~,  \notag \\
\mathbf{\Omega }_{\mathbf{i}}^{SS} &=&\frac{GS_{j}}{c^{2}r^{3}}\left[ 3%
\mathbf{\left( \mathbf{\hat{r}}\cdot \hat{S}_{j}\right) \hat{r}-\hat{S}}_{%
\mathbf{j}}\right] ~,  \notag \\
\mathbf{\Omega }_{\mathbf{i}}^{QM} &=&-\frac{3Gm_{j}Q_{i}}{r^{3}S_{i}}%
\mathbf{\left( \mathbf{\hat{r}}\cdot \hat{S}_{i}\right) \hat{r}~,}
\label{Omegas}
\end{eqnarray}%
where $j\neq i$. The sum of the SS and QM contributions, by employing Eqs. (%
\ref{mi})-(\ref{Qi}) is 
\begin{equation}
\mathbf{\Omega }_{\mathbf{i}}^{SS}+\mathbf{\Omega }_{\mathbf{i}}^{QM}=\frac{G%
}{c^{2}r^{3}}\left[ 3\mathbf{\left( \mathbf{\hat{r}}\cdot S_{j}\right) \hat{r%
}}+3w\nu ^{2\left( j-i\right) }\mathbf{\mathbf{\left( \mathbf{\hat{r}}\cdot
S_{i}\right) \hat{r}}-S}_{\mathbf{j}}\right] ~,\quad \mathbf{j}\neq \mathbf{i%
}.  \label{OmSSQM}
\end{equation}

In order to evaluate the PN\ order of the coefficients in Eqs. (\ref{Omegas}%
), we will employ the estimate from a footnote of Ref. \cite{spinspin2},
according to which%
\begin{equation}
\mathcal{O}\left( \frac{c}{r}\right) =\varepsilon ^{-1/2}\mathcal{O}%
(T^{-1})~,
\end{equation}%
$T$ being the radial period, defined as twice the time elapsed between
consecutive $\dot{r}=0$ configurations. We obtain%
\begin{eqnarray}
{\frac{G\left( 4+3\nu ^{3-2i}\right) }{2c^{2}r^{3}}L}_{N} &=&\frac{1}{2}{%
\frac{Gm}{c^{2}r}}\frac{v}{c}\frac{c}{r}\frac{{L}_{N}}{mrv}\left( 4+3\nu
^{3-2i}\right) =\mathcal{O}\left( \varepsilon \right) \mathcal{O}\left( \nu
^{-1},1,\nu \right) \mathcal{O}(T^{-1})~,  \notag \\
\frac{GS_{j}}{c^{2}r^{3}} &=&\left( \frac{Gm}{c^{2}r}\right) ^{2}\frac{c}{r}%
\eta \nu ^{2i-3}\chi _{j}=\mathcal{O}\left( \varepsilon ^{3/2}\right) 
\mathcal{O}\left( \nu ^{-1},\nu \right) \chi _{j}\mathcal{O}(T^{-1})~, 
\notag \\
-\frac{3Gm_{j}Q_{i}}{r^{3}S_{i}} &=&3w\left( \frac{Gm}{c^{2}r}\right) ^{2}%
\frac{c}{r}\eta \chi _{i}=\mathcal{O}\left( \varepsilon ^{3/2}\right) 
\mathcal{O}\left( \eta \right) \chi _{i}\mathcal{O}(T^{-1})~.
\label{LS1S2estimates}
\end{eqnarray}%
Thus on the orbital time-scale the SO precession is a 1PN effect, while the
SS and QM\ contributions appear as 1.5 PN corrections. As both the SO and SS
angular velocities contain terms with $\mathcal{O}\left( \nu ^{-1}\right) $
factors, whenever the mass ratio is small, the respective precessions
amplify.

As $\mathbf{\Omega }_{i}^{QM}\propto \chi _{i}$, the QM precession qualifies
as a \textit{self-spin} effect.

\subsection{Spin configurations preserved by precessions}

With only the leading order SO precession taken into account, both spin
vectors undergo a precession about $\mathbf{\hat{L}_{N}}$. If $m_{2}=m_{1}$
also holds, the instantaneous angular velocities of the precessions are
identical, and the spin configuration is preserved with respect to the
osculating plane of the orbit, rigidly rotating about its normal.

With the SS and QM contributions to the spin dynamics included, the above
simple picture does not hold any more. In the remaining part of this section
we analyze whether there are spin configurations which are preserved by
precessions, in the sense that they rigidly precess about a common rotation
axis.

We will carry on this analysis order by order, starting with the leading
order SO precession. One possibility is that both spins are either aligned
or antialigned with the orbital angular momentum $\mathbf{\hat{S}_{i}=\pm 
\hat{L}_{N}}$, then there is no precession at SO order. Moreover, at the
next order we immediately obtain $\mathbf{\Omega }_{\mathbf{i}}^{QM}=0$ and $%
\mathbf{\Omega }_{\mathbf{i}}^{SS}\propto \mathbf{\hat{L}_{N}}$, such that $%
\mathbf{\dot{S}_{i}}=0$. Thus, when the spins are perpendicular to the
osculating orbit at some initial instant, they stay so, even with the SS and
QM\ parts of the dynamics included.

Another possibility to consider is, that the two spins precess with the same
angular velocity about a common axis. We could check, whether the axis
defined by Eq. (\ref{Omegas}) could be this, however we will allow for more
generic possibilities. As $\mathbf{S}_{\mathbf{i}}$ undergo pure
precessions, one can add arbitrary contributions $\left( G/c^{2}r^{3}\right)
\left( \mathcal{P}_{i}-1\right) \mathbf{S}_{\mathbf{i}}$\ to $\mathbf{\Omega 
}_{\mathbf{i}}$ without changing the dynamics, and ask the question, whether
a common instantaneous axis of precession exists for both spin vectors,
about which they precess with equal angular velocities, such that $\mathbf{%
\Omega }_{\mathbf{1}}^{\prime }=\mathbf{\Omega }_{\mathbf{2}}^{\prime }$?
The condition for this would be%
\begin{equation}
0={\frac{\left( \nu -\nu ^{-1}\right) }{2}}\mathbf{L}_{\mathbf{N}}+\left\{ 
\mathbf{\hat{r}}\cdot \left[ \left( 1-w\nu ^{-1}\right) \mathbf{S}_{\mathbf{2%
}}-\left( 1-w\nu \right) \mathbf{S}_{\mathbf{1}}\right] \right\} \mathbf{%
\hat{r}-}\frac{1}{3}\left( \mathcal{P}_{2}\mathbf{S}_{\mathbf{2}}-\mathcal{P}%
_{1}\mathbf{S}_{\mathbf{1}}\right) ~.  \label{diffOm}
\end{equation}%
For $\mathcal{P}_{i}$ of order unity (meaning that this axis is not very far
from the normal to the osculating orbit) the leading order contribution in
Eq. (\ref{diffOm}) remains the term proportional to $\mathbf{L}_{\mathbf{N}}$%
, the vanishing of which implies $m_{2}=m_{1}$. For the next order then we
get 
\begin{equation}
\mathcal{P}_{2}\mathbf{S}_{\mathbf{2}}-\mathcal{P}_{1}\mathbf{S}_{\mathbf{1}%
}=3\left( 1-w\right) \left[ \mathbf{\hat{r}}\cdot \left( \mathbf{S}_{\mathbf{%
2}}-\mathbf{S}_{\mathbf{1}}\right) \right] \mathbf{\hat{r}}~.
\label{diffOm1}
\end{equation}%
For black holes ($w=1$) this gives $\mathcal{P}_{2}\mathbf{S}_{\mathbf{2}}=%
\mathcal{P}_{1}\mathbf{S}_{\mathbf{1}}$, thus \textit{the spins should be
parallel} (aligned or antialigned), and the common axis of synchronous
rotation is 
\begin{equation}
\mathbf{\Omega }_{\mathbf{i}}^{\prime }={\frac{G}{c^{2}r^{3}}}\left[ {\frac{7%
}{2}}\mathbf{L_{N}}+3\mathbf{\left( \mathbf{\hat{r}}\cdot S\right) \hat{r}-S}%
+\frac{\mathcal{P}_{1}\mathbf{S}_{\mathbf{1}}+\mathcal{P}_{2}\mathbf{S}_{%
\mathbf{2}}}{2}\right] ~,  \label{Om1}
\end{equation}%
with $\mathbf{S}=\mathbf{S}_{\mathbf{1}}+\mathbf{S}_{\mathbf{2}}$. Neither
the axis of rotation nor the angular velocity are unambiguous, as $\mathbf{%
\Omega }_{\mathbf{i}}^{\prime }$ depend on $\mathcal{P}_{i}$, however the
axis stays close to $\mathbf{\hat{L}_{N}}$ (no choice of $\mathcal{P}_{i}$
would render the axis of rotation exactly to $\mathbf{\hat{L}_{N}}$). In
summary, only parallel black hole spins can rotate with the same angular
velocity about a common axis, provided the axis is only slightly different
from the normal to the osculating orbit.

\section{Concluding Remarks}

In this paper we have derived the set of independent variables suitable to
monitor the evolution of a compact spinning binary during the inspiral. The
number of independent variables characterizing the spins and orbital angular
momentum was shown to be 6. We have chosen them either as 5 angles and a
scale, or alternatively as 3 angles and 3 scales. For the first choice we
found advantageous to employ the magnitude $J$\ of the total angular
momentum; the angles $\alpha $ and $\kappa _{i}$\ span by the Newtonian
orbital angular momentum $\mathbf{L}_{\mathbf{N}}$ with the total angular
momentum $\mathbf{J}$ and with the spins, respectively; finally the
azimuthal angles $\psi _{i}$ of the spins in the plane of motion
(perpendicular to $\mathbf{L}_{\mathbf{N}}$), measured from the ascending
part of the node line (the intersection of the planes perpendicular to $%
\mathbf{J}$\ and $\mathbf{L}_{\mathbf{N}}$.) For the second choice we
propose $J$, $\alpha $, $\kappa _{i}$ and the normalized magnitudes of the
spins $\chi _{i}$. As both $J$ and $\chi _{i}$ are unaffected by
precessions; moreover $\chi _{i}$ vary extremely slowly with gravitational
radiation reaction, the latter set seems more advantageous. Nevertheless,
expressing $\psi _{i}$ in the detriment of $\chi _{i}$ is not immediate (the
respective equations are provided).

These 6 variables have to be supplemented by the true anomaly $\chi _{p}$.
The non-inertial character of the reference systems introduced in Section 3
can be specified through one single angle $\phi _{n}$, characterizing the
node line, the evolution of which is governed by the spin-orbit coupling.
The orbital evolution being quasi-Keplerian, the position of the periastron
is specified by an evolving angle $\psi _{p}$. As shown in subsection \ref%
{EulerConstr}, the evolution of these two angles follow from the evolution
of $\alpha $ and $\chi _{p}$.

In this paper we have also proven a no-go result, according to which in a
2PN accurate dynamics, with the leading order SO, SS and QM precessions
included the only binary black hole configuration allowing for spin
precessions with equal angular velocities about a common instantaneous axis
roughly aligned to the normal to the osculating orbit, is the \textit{equal
mass and parallel (aligned or antialigned) spin }configuration\textit{.}
When including only the SO precessions, the equality of masses is required,
but there is no constraint on the spin orientations. By approaching the
innermost stable orbit, the PN parameter increases (leading eventually to
the breakdown of the PN expansion), such that the importance of higher order
contributions is enhanced. Therefore the SS and QM precessions (of higher
order than the SO precession), which lead to the above constraint on the
spin directions, become increasingly larger. The result thus will hold up to
the very last orbits of the inspiral, and to the extent the PN result
approximates well dynamics there, during the plunge. This analytic result
puts limitations on what particular precessing configurations can be
selected in numerical\ investigations of compact binary evolutions, even in
those including only the last orbits of the inspiral.

\section{Acknowledgements}

I acknowledge stimulating discussions with Zolt\'{a}n Keresztes. This work
was supported by the Hungarian Scientific Research Fund (OTKA) grant no.
69036 and the Pol\'{a}nyi Program of the Hungarian National Office for
Research and Technology (NKTH).

\end{document}